\documentclass{aa}
\usepackage{graphics}

\begin{document}      

   \title{A method for determining radio continuum spectra, and its application to large surveys}

   \author{B.~Vollmer\inst{1,2}, E.~Davoust\inst{3}, P.~Dubois\inst{1}, F.~Genova\inst{1}, F.~Ochsenbein\inst{1}, W.~van~Driel\inst{4}}

   \offprints{B.~Vollmer, e-mail: bvollmer@astro.u-strasbg.fr}

   \institute{CDS, Observatoire astronomique de Strasbourg, UMR 7550, 11, rue de l'universit\'e, 
     67000 Strasbourg, France \and
     Max-Planck-Institut f\"ur Radioastronomie, Auf dem H\"ugel 69, 53121 Bonn Germany \and
     UMR 5572, Observatoire Midi-Pyr\'{e}n\'{e}es, 14 avenue E. Belin, 31400 Toulouse, France \and
     Observatoire de Paris, Section de Meudon, GEPI, CNRS UMR 8111 and Universit\'e Paris 7, 
     5 place Jules Janssen, 92195 Meudon Cedex, France
   } 
          
   \date{Received / Accepted}

   \authorrunning{Vollmer et al.}
   \titlerunning{SPECFIND}

\abstract{
A new tool to extract cross-identifications and radio continuum spectra from radio catalogues 
contained in the VIZIER database of the CDS is presented. 
The code can handle radio surveys at different frequencies with different resolutions.
It has been applied to 22 survey catalogues at 11 different frequencies containing a total of 3.5
million sources, which resulted in over $700\,000$ independent radio cross-identifications
and $\sim 67\,000$ independent radio spectra with more than two frequency points.
A validation of the code has been performed using independent radio cross-correlations from the literature. 
The mean error of the determined spectral index is $\pm 0.3$.  The code produces an output 
of variable format that can easily be adapted to the purpose of the user. 
}

\maketitle

\section{Introduction \label{sec:intro}}

The total number of records in radio-source catalogues has increased 
dramatically in the last two decades. Three major increases are due to
R.~Dixon's 'Master Source List' in the seventies 
(for the first version see Dixon 1970;
for an error list see Andernach 1989), to the 87GB (Gregory \& Condon 1991),
GB6 (Gregory et al. 1996), and PMN (Wright et al. 1994, 1996;
Griffith et al. 1994, 1995) surveys in 1991, and to the
release of NVSS (Condon et al. 1998), FIRST (White et al. 1998) and WENSS 
(Rengelink et al. 1997) in 1996/1997 (see also Andernach 1999).
In order to study the nature of these celestial objects detected at radio frequencies 
and to take full advantage of the huge amount of information contained in the catalogues, 
one has to know their spectral energy distribution (SED) over as large a frequency
range as available. Since each catalogue (except Dixon's master source list)
contains information at a single frequency,
cross-identification between different catalogues is essential for the
study of these sources. 

Radio source cross-identifications in the centimeter to meter wavelength domain 
are particularly difficult to obtain, because the underlying radio surveys
can have huge differences in sensitivity and/or spatial resolution.
Since the resolution depends on the observed frequency and the
telescope diameter, low frequency surveys
made with a small single-dish telescope can have resolutions up to
tens of arcmins, while high-frequency observations with a large single
dish telescope or an interferometer can have resolutions of a few arcsecs 
to tens of arcsecs. 

On the other hand, the cross-identification of radio sources
at different frequencies is made easier by using the fact that, in the vast majority of sources, the
SED has a power-law distribution.  The radiation mechanism  is either synchrotron emission from relativistic
electrons gyrating in a magnetic field, or emission of hot thermal electrons. 
Synchrotron emission produces a power law spectrum with a possible cut-off
or reversal of the spectral index at low frequencies due to self-absorption or 
comptonization. The spectrum of thermal electrons is flat, at least in the optically thin domain. 
Over the frequency range in which the majority of radio surveys were made, 
the spectra are thus well defined by a power law, i.e. as a straight line
in the $\log$(flux density)--$\log$(frequency) plot commonly used in radio continuum astronomy. 

The cross-identification procedure assigns radio sources at different frequencies to one 
physical object. In this way accurate radio positions can be determined for
these objects using the high frequency observations, their radiation processes can be studied, 
and a search for specific objects (data mining) becomes possible. 

The VIZIER database\footnote{http://vizier.u-strasbg.fr/viz-bin/VizieR} (Ochsenbein et al. 2000)
at the Centre de Donn\'ees astronomique de Strasbourg (CDS)
contains approximately 500 catalogues with radio data. Of these
500, about 70 catalogues are from systematic surveys (for a list of the major surveys see Andernach 1999).
Using VIZIER, only a cone search (where a central position and a radius is used) is possible on 
all, or on a subset of, these catalogues. This procedure gives a list of all radio
sources within the search region without any cross-identification.

Cross-identifications of radio sources within SIMBAD\footnote{http://simbad.u-strasbg.fr/sim-fid.pl} 
(Wenger et al. 2000) are essentially made on bibliographic grounds. Sources are only merged when a newly published
radio catalogue gives alternate names, which were already known to SIMBAD. Thus there is a clear need
for establishing links between the radio catalogues (cross-identifications) and for including 
the sources into SIMBAD. 

\subsection{Radio surveys}

There are two main kinds of radio surveys: (i) single dish and (ii) interferometric.
Since interferometers cannot detect extended structures larger than the 
size corresponding to the angular resolution  of the shortest baselines (ten/tens of arcmin for
compact configurations), only structures
of this size, or smaller, can be detected in sources of large extent. This makes the identification and unique spectral 
index determination complex for very extended sources, if they were observed with both single-dish 
and interferometer telescopes.

In addition, the observations can be divided into two subclasses: (i)systematic surveys and
(ii) surveys of a given source sample (e.g. SN remnants, H{\sc ii} regions, etc.).
While the first category is more suitable for integration into VIZIER,
the second one is easier to feed into SIMBAD.
Out of the 500 catalogues listed in VIZIER when choosing ``radio'', about 220 contain
independent observations. Of these, about 80 represent systematic surveys and 
about 140 surveys of source samples.

The characteristics of radio surveys that have to be taken into account when searching for 
cross-identifications are 
\begin{itemize}
\item
frequency,
\item
sky coverage,
\item
angular resolution (half power beam width, HPBW),
\item
sensitivity to point-like and extended sources.
\end{itemize}

From these surveys source catalogues are established.
The sources are extracted from 2D maps using 2D Gaussian fits, which have in
principle 4 parameters: (i) the center position, (ii) the major axis, 
(iii) the minor axis, and (iv) the position angle.
One distinguishes 'map' parameters and 'sky' parameters of sources. 'Map'
parameters are the extent along the major and minor axis from the Gaussian fit,
whereas 'sky' parameters are the deconvolved ones. 
If $\Theta$ is the survey resolution (HPBW) and
$d$ the source extent on the map, then the true source extent is 
$d_{\rm true} = \sqrt{d^{2}-\Theta^{2}}$. 

The catalogues can have different formats.
For a given source one may find positions, position errors, a name, the peak 
and integrated 
flux densities and their errors, the major/minor axis, the position angle, and various flags
(confusion, extended source, warnings, etc.). The use of unified content descriptors 
(UCDs\footnote{see http://vizier.u-strasbg.fr/doc/UCD/inVizieR.htx}),
which is a classification scheme in which all the astronomical parameters accessible in VIZIER are stored,
will make the data access much easier in the future.

\subsection{Cross-identification of radio sources}

There are three categories of cross-identification between radio sources:
\begin{itemize}
\item
cross-identification based on the proximity between two sources,
\item
cross-identification based on the radio spectrum,
\item
cross-identification based on other physical characteristics (e.g. galaxies, SN remnants, 
quasars, AGNs etc.). 
\end{itemize}
The last two kinds of cross-identifications are called 'value-added'
cross-identification.
In the present paper we present a tool for cross-identification called SPECFIND
which uses a proximity criterion and assumes that all sources have
a power law radio spectrum.

This paper is structured in the following way: Sect.~\ref{sec:sql} describes
the uniformisation of catalogue tables, which is 
required because the catalogue table entries often differ.
The code algorithm is discussed in Sect.~\ref{sec:general},
followed by a discussion on the detailed code 
structure (Sect.~\ref{sec:detailed}). The code performance is presented in 
Sect.~\ref{sec:performance} and the results are shown in Sect.~\ref{sec:results}.
The code is validated by the comparison of our spectral indices with independent 
estimates from the literature (Sect.~\ref{sec:validation}).
The summary and conclusions are given in Sect.~\ref{sec:conclusions}.

\section{Preparation and uniformisation of radio catalogues \label{sec:sql}}

The main difficulty in treating simultaneously different radio catalogue tables
lies in the variety of their table structures. All tables have entries for the
name, coordinates, and the flux density of the sources. The name and coordinates
can be in different epochs (B1950 or J2000). Other possible entries are:
\begin{itemize}
\item
position errors in right ascension and declination (including pointing errors and
the uncertainty with which the center of the 2D Gaussian could be determined),
\item
flux density error,
\item
major/minor axis diameter of the source of the fitted ellipse (not deconvolved),
\item
position angle of the fitted ellipse,
\item
different sorts of flags (warning, border of the observed field, possible confusion, interferences, etc.).
\end{itemize}
This variety of entries makes a uniformisation of the radio catalogues unavoidable.
 
SPECFIND uses its own standard for the catalogue entries. All coordinates
are in J2000 and the source names are chosen to be in accordance with the SIMBAD
nomenclature. Table~\ref{tab:standard} shows the list of SPECFIND catalogue entries.
\begin{table}	
      \caption{SPECFIND catalogue standard.}
         \label{tab:standard}
      \[
         \begin{array}{llll}
{\rm RA (2000)}    &	{\rm arcsec}		&	&	{\rm REAL}         \\
{\rm RA\_error}    &	{\rm arcsec}		&	&	{\rm REAL}         \\ 
{\rm DEC(2000)}    &	{\rm arcsec}		&	&	{\rm REAL}         \\
{\rm DEC\_error}   &	{\rm arcsec}		&	&	{\rm REAL}         \\
{\rm Flux}	   &    {\rm mJy}		&	&	{\rm REAL}         \\
{\rm flux\_error}  &	{\rm mJy}	        &	&	{\rm REAL}         \\
{\rm Wflag}	   &	{\rm warning}		& 0/1	&	{\rm INT}      \\
{\rm Cflag}	   &	{\rm confused\ source}   & 0/1	&	{\rm INT}      \\
{\rm Eflag}	   &	{\rm extended\  source}  & 0/1	&	{\rm INT}      \\ 
{\rm MajAxis}	   &	{\rm major\ axis\ (units\ of \ beamsize)} & & {\rm REAL}         \\
{\rm MinAxis}	   &	{\rm minor\ axis\ (units\ of \ beamsize)} & & {\rm REAL}         \\
{\rm PA}	   &	{\rm position\ angle\ (degrees)} &  &	{\rm REAL}         \\
{\rm name2000}     &	{\rm source\ name\ J2000}	&	&	{\rm CHAR}  \\ 
{\rm name1950}     &	{\rm source\ name\ B1950}	&	&	{\rm CHAR}  \\
\end{array}
      \]
\end{table}
The integrated flux densities $S_{\rm int}$ are taken directly from the radio catalogues,
except for the GB6, 87GB, and MIYUN catalogues, which give only
peak flux densities $S_{\rm peak}$. In the GB6 and 87GB catalogues we take 
the peak flux density as the integrated flux density for sources smaller than
1.1 times the beamsize (3\farcm5), and calculate the integrated flux density as
\begin{equation}
S_{\rm int}=S_{\rm peak} \big( \frac{d_{\rm maj}}{\Theta} \big) \big( \frac{d_{\rm min}}{\Theta}\ \big),
\end{equation}
for larger sources, where $d_{\rm maj}$ and $d_{\rm min}$ are the extents along the major and minor axis
on the map, respectively. 
For the MIYUN catalogue we take the peak flux density as integrated flux density
when the latter is not given explicitly.
The values of the flux errors that are not directly taken from the catalogues
are listed in Table~\ref{tab:fluxerror}.
\begin{table}	
      \caption{Definitions of the flux density error when not taken directly from the catalogues.}
         \label{tab:fluxerror}
      \[
         \begin{array}{ll}
	{\rm JVAS} & 0.1\times{\rm Flux} \\
	{\rm BWE}  & \sqrt{36.+(0.01\times{\rm Flux}^{2})} \\
	{\rm PKS}  & {\rm Flux/SN} \\
	{\rm PKS}  & 0.2\times{\rm Flux} \\
	{\rm F3R}  & {\rm error\ code} \\
	{\rm B2}   & 0.2\times{\rm Flux} \\
	{\rm B3}   & 0.2\times{\rm Flux} \\
	{\rm TXS}  & 0.2\times{\rm Flux} \\
	{\rm WISH} & {\rm flux\_error\ or\ }0.15\times{\rm Flux} \\
	{\rm WENSS\ \ \ \ \ }& {\rm flux\_error\ or\ }0.15\times{\rm Flux} \\
	{\rm MIYUN\ \ \ \ \ }& {\rm peak\ flux\ or\ Flux/SN} \\
	{\rm 4C}   & 0.3\times{\rm Flux} \\
	{\rm 4C}   & 0.3\times{\rm Flux} \\
\end{array}
      \]
\end{table}

The flags are based on those taken from the catalogues.
For the moment, however, SPECFIND does not make use of the different flags.

\section{SPECFIND -- the algorithm \label{sec:general}}

SPECFIND is a hierarchical code. It classifies a source $j$ as parent, sibling, or
child with respect to a given source $i$ at different stages where stage 2 and 3 are
refinements of stage 1.
\begin{itemize}
\item
stage 1: depending on proximity criteria: 
\begin{itemize}
\item
parent: source $j$ has a larger extent or was observed with a lower angular 
resolution than source $i$,
\item
sibling: source $j$ has a comparable extent or was observed with a comparable 
angular resolution (within 25\%) to that of source $i$,
\item
child: source $j$ has a smaller extent or was observed with a higher angular 
resolution than source $i$.
\end{itemize}
\item
stage 2: depending on flux densities at the same frequency:
\begin{itemize}
\item
parent: source $j$ has a larger extent or resolution and has a larger flux density 
than source $i$,
\item
sibling: source $j$ has a comparable extent or resolution and has the same flux density
within the errors as source $i$,
\item
child: source $j$ has a smaller extent or resolution and a smaller flux density
than source $i$.
\end{itemize}
\item
stage 3: depending on flux densities at different frequencies, based on the expected radio
spectral index:
\begin{itemize}
\item
parent: source $j$ has a larger flux density than expected from the radio spectrum that 
includes source $i$,
\item
sibling: source $j$ fits into the radio spectrum that includes source $i$,
\item
child: source $j$ has a smaller flux density than expected from the radio spectrum that 
includes source $i$.
\end{itemize}
\end{itemize}
At the end of this procedure source
 $i$ and its siblings are considered the same source.
If there are parents, source $i$ might be a resolved sub-source of source $j$.
If there are children, source $i$ might be extended without resolving the children,
and the children represent the sub-sources of source $i$.
SPECFIND also adds the flux densities of the children of source $i$ at the same frequency.
If the sum equals the flux density of source $i$ within the errors, then source $i$ is
considered as extended without resolving the children.

While the parents and children at the same frequency are not used for the moment,
the parents and children at different frequencies are taken together with the
siblings to determine the radio spectrum.

\section{SPECFIND -- detailed code structure \label{sec:detailed}}

\subsection{Proximity search \label{sec:prox}}

The proximity search is done using the treecode-method routines written by J.~Barnes
(see e.g. Barnes \& Hut 1986 
\footnote{http://www.ifa.hawaii.edu/$\sim$barnes/software.html}). Since this code is well
documented, we will not describe the method nor the routines.
We adapted these routines to spherical geometry. 
The angular distance between two sources, which are located at $(\alpha_{1}, \delta_{1})$ and
$(\alpha_{2}, \delta_{2})$ is calculated in spherical geometry:
\begin{equation}
d=\arccos ( \cos(\alpha_{1}-\alpha_{2}) \cos (\delta_{1}) \cos (\delta_{2}) +\sin(\delta_{1}) \sin(\delta_{2}))\ . \label{sphericald}
\end{equation}
We included the possibility to check if source $j$ is located
within the Gaussian ellipsoid characterising source $i$.
If this is the case, the separation of the sources into parents/siblings/children is done in the following way:
let $d_i$=max(resolution($i$),Majaxis($i$)) and $d_j$=max(resolution($j$),Majaxis($j$)), i.e.
the maximum between the resolution with which source $i$/$j$ was observed and
their major axes. If $d_j > 1.25\,d_i$, then source $j$ is considered as a parent.
If $0.75\,d_i \le d_j \le 1.25\,d_i$, source $j$ is considered as a sibling.
If $d_j < 0.75\,d_i$, source $j$ is considered as a child.
We do not include a check for the positional error ellipsoid, because when they are given in a catalogue
these are only small fractions of the beamsize. On the other hand, since we take the FWHM of the Gaussian
fit as source extent, it is not necessary to take into account the error ellipsoid of the fit.

Since the treecode works in plane geometry, the polar caps ($|\delta| > 70^{\circ}$) 
have to be treated separately. The rest of
the sky is divided into equal RA slices, which ensures an approximately equal number
of sources per slice. For the next neighbour search within each slice, sources within a
somewhat larger field than the slice are used. 
The RA offset between this field and the slice is taken to be three times
the largest source extent of all catalogues.\\
Since the largest extent of all sources is $1^{\circ}$, 
%and thus
the maximum overlap is $3^{\circ}$.
At a declination of  $70^{\circ}$ a linear separation in RA of $(\alpha_{1}-\alpha_{2})=3^{\circ}$ 
corresponds to an angular separation of $1^{\circ}$. 
Proximity searches within both polar regions were also performed without using the treecode
($N \times N$ calculations), which showed no change of the result with respect to the treecode.\\
For checking if source $j$ is located within the ellipsoidal source extent of source $i$,
the ARC\footnote{see Greisen (1994)\\ http://www.aoc.nrao.edu/aips/aipsmemo.html} 
projection is used for all catalogues. This projection, which preserves angular distances, is used
for Schmidt plates (to first order) and in mapping with single-dish radio telescopes.
Although the SIN$^{4}$ projection is applied for interferometric maps, 
we used the ARC projection for these as well, since (i) the
majority of our catalogues are single-dish measurements, (ii) the largest source extents
are found in single dish observations because of their larger beamsize, and (iii) the ARC 
and SIN projections are similar.
An additional search is included for sources that are located around RA = 00$^{\rm h}$00$^{\rm m}$00$^{\rm s}$.

\subsection{Joining sources at the same frequency \label{sec:join}}

This routine takes into account the flux densities of the sources at the same frequency.
Since the sources are frequently point-like, i.e. their size fitted from the map equals 
the beamsize/resolution of the antenna used (unresolved sources), 
a source observed with a given beamsize is classified as a parent of another source observed 
with a smaller beamsize at the same frequency. Consequently, a source observed with a given beamsize
is classified as a child of another source observed with a larger beamsize at the same frequency. 
A source $i$ can have only one parent but several children from a different survey at the same frequency.
Some or all children may  be the same physical object 
-- this can be verified with the help of the flux densities measured at the same frequency.
The routine makes the following classification:
a parent has a larger flux density and a larger extent/resolution,
a sibling has an equal flux density within the errors and an equal extent/resolution within 25\%,
a child has a smaller flux density and a smaller extent/resolution.

\subsection{Check for family dependences}

Once the sources are joined in the way described in Sect.~\ref{sec:join},
the family dependences are verified, i.e. for a given source cross-checks are performed.
These checks are performed for all source entries;
\begin{itemize}
\item
If source $j$ is a sibling of source $i$, then source $i$ must also
be a sibling of source $j$.
\item
If source $j$ is a child of source $i$, source $i$ must be a parent of source $j$.
\item
If source $j$ is a parent of source $i$, source $i$ must be a child of source $j$.
\end{itemize}
In addition, the flux densities of all children are added to investigate 
whether the source
has multiple components at a given frequency. If the sum of the flux densities
of all siblings equals that of the source within the errors,
it is considered as resolved and an internal flag is set, which tells the user at which frequency
the source is resolved. 
It happens frequently that a non-Gaussian emission distribution
consists of multiple components within the area of one beamsize when observed with a smaller beam.

\subsection{Spectrum-finding algorithm \label{sec:spectra}}

This is the most important routine and thus the heart of SPECFIND.
It uses the method of the least absolute deviation to make a linear
fit in the $\log \nu$--$\log S_{\nu}$ plane, where $\nu$ is the frequency and $S_{\nu}$
is the flux density at frequency $\nu$. This method is more robust against
outlying points in a spectrum than a standard least-squares deviation ($\chi^{2}$) fit (see Press et al. 2002).

For this algorithm, the best way to find a maximum number of spectra without a too
high risk of misidentifications 
is to set 
%was found to be setting 
the flux density errors of all sources that are smaller 
than 20\% of their flux density to this 20\% value and multiplying flux density errors by
a factor 1.5. In this way all catalogues have approximately the same
relative error. Moreover, these relatively large errors can compensate a not too
strong flattening of the spectral index at low frequencies due to the
synchrotron turnover or at high frequencies due
to an increasing fraction of thermal emission.

The structure of the spectrum-finding algorithm is the following:
for a given set of sources for which all family relations were determined,
their flux measurements at different frequencies are grouped together into 
an array and sorted by frequency. If the number of different frequencies is greater than
two, the spectrum-finding algorithm passes through the following steps:
\begin{enumerate}
\item
a least absolute deviation fit in the $\log S_{\nu} - \log \nu$ plane is performed:
\begin{equation}
\log S_{\nu}=\alpha \log \nu + \gamma\ ;
\label{eq:spec}
\end{equation}
\item
if the spectrum is determined more than once:\\
if the number of sources that fit into the spectrum decreases, the old
fit parameters are used;
\item
check for sources that fit into the spectrum; if all sources fit, go to step 6.;
\item
if there are two sources of the same frequency, the one with the largest
deviation from the fit is flagged and removed;
\item
if all sources have different frequencies the source with the largest
deviation from the fit is flagged and removed, go to step 1.;
\item
if there are more than two independent points left and if
the ratio between the largest and the smallest frequency interval is
greater than 0.02, make a final fit;
\item
go to step 1. and make a second run with $\alpha=-0.9$ and $\gamma=\log S_{\nu}-\alpha \log \nu$
during the first $N-4$ steps of the loop, where $N$ is the initial number of points
in the spectrum (-- 0.9 is the mean spectral index of all radio sources);
\item
if the number of fitted points with fixed $\gamma$ and $\alpha$ exceeds that of the
initial fitting procedure, this spectrum is accepted; otherwise the
spectrum of the first fitting procedure is accepted.
\end{enumerate}
Step 4 excludes variable sources that are observed at different epochs. 
If the majority of the points has
a high angular resolution, extended sources that are resolved into
sub-sources might also be discarded.
This algorithm turned out to be the most promising for finding
a maximum number of spectra with only a small risk of misidentifications
(see Sect.~\ref{sec:results}).
By definition it can only detect parts of a spectrum that follow a power law
(see also Sect.~\ref{sec:complete}).
This algorithm is similar to that used by Verkhodanov et al. (2000) for the
identification of radio spectra of decameter-wavelength sources.

In order to investigate the efficiency of the spectrum-finding algorithm,
we inspected by eye the data of the sources for which no spectrum could be 
found. 
We optimised the code to find a maximum number of radio spectra with a relatively 
small number of misidentifications (see also Sect.~\ref{sec:validation}).

\subsection{Check for frequency intervals and ambiguous sources}

In order to avoid using points that are too close to each other in frequency,
and therefore not independent, the frequency intervals between the different
points of the spectrum are checked.
The routine calculates the frequency intervals and determines 
the ratio between the second largest and the largest frequency interval.
If this ratio is smaller than 0.02, the spectrum is rejected.

Then, in order to avoid ambiguous radio sources of a given frequency, which are 
attributed to two distinct physical objects, the ``center of mass'' coordinates
are calculated for both objects, where the inverse of the survey resolution is used 
for the ``mass''. The ambiguous source is then attributed solely to the object
whose ``center of mass'' position is nearest to the source position.

\subsection{Completeness and uniqueness check for spectra \label{sec:complete}}

This routine insures that if a source $j$ fits the spectrum determined for 
source $i$ (where source $i$ is included), then source $i$ also appears in
the spectrum of source $j$. In this way it is insured that a radio source
belongs to only one single physical object. 

In practice the spectrum-finding algorithm is too efficient. Through chance alignments,
sources that were observed with a large beamsize or which have a large extent
are sometimes identified as physically belonging together (via the radio spectrum).
Thus it happens that, while a source $j$ fits the spectrum determined for 
source $i$, source $i$ is not included into the spectrum of source $j$.
There may be several reasons for this: (i) the spectrum of source $i$ is erroneous,
(ii) the spectrum of source $j$ is erroneous, (iii) the spectral
index varies with frequency, (iv) the real errors on the
flux densities of one of the sources are larger than $30\%$. 
The task to make all spectra consistent is quite complicated, because all
sources are interconnected via their siblings, which are interconnected
via their own siblings, etc. 

In order to decide which spectrum to take, in case of inconsistency between two
spectra, the following scheme is applied:\\
let $\alpha_{i},\ \alpha_{j}$ be the spectral indices of source $i$ and $j$.
(i) If the difference between the spectral indices is smaller than 0.3 $(|\alpha_{i}-\alpha_{j}|<0.3)$ 
(see Sect.~\ref{sec:validation}) both spectra are real.\\
(ii) If $(|\alpha_{i}-\alpha_{j}| \geq 0.3)$ and
the frequency intervals within which they are determined are only marginally
overlapping (20\%) both spectra are real. The spectrum is thus approximated by two different slopes
within two different frequency intervals.\\
(iii) if (i) and (ii) are not the case, the spectrum with the larger number of independent
frequency points is real. The other spectrum is discarded.\\
(iv) if (i) and (ii) are not the case and numbers of independent frequency points
are the same for both spectra:\\
(iva) if $\alpha_{i} < 0$ and $\alpha_{j} < 0$ and
$\alpha_{i} > \alpha_{j}$, spectrum $i$ is real and spectrum $j$ is discarded;\\
(ivb)  if $\alpha_{i} < 0$ and $\alpha_{j} < 0$ and
$\alpha_{i} < \alpha_{j}$, spectrum $j$ is real and spectrum $i$ is discarded;\\
(ivc) if $\alpha_{i} \times \alpha_{j} < 0$ and $\alpha_{i} < 0$, spectrum $i$ is real and 
spectrum $j$ is discarded;\\
(ivd) if $\alpha_{i} \times \alpha_{j} < 0$ and $\alpha_{j} < 0$,  
spectrum $j$ is real and spectrum $i$ is discarded;\\ 
(ive) if $\alpha_{i} > 0$ and $\alpha_{j} > 0$ and
$\alpha_{i} < \alpha_{j}$, spectrum $i$ is real and spectrum $j$ is discarded;\\
(ivf)  if $\alpha_{i} > 0$ and $\alpha_{j} > 0$ and
$\alpha_{i} > \alpha_{j}$, spectrum $j$ is real and spectrum $i$ is discarded.
The spectra are only completed if the proximity criterion is satisfied,
i.e. if the extents/beamsizes of the sources intersect.

The above described procedure checks only for siblings of siblings. This does not
take into account the case where the same source is attached as a sibling to two
different sources without being a sibling of a sibling of any of these two sources.
The only way to identify these cases is to make again a full proximity search using
the treecode (Sect.~\ref{sec:prox}). All nearby sources are then checked for  
common siblings. If the modulus of the difference between the spectral indices of the two
sources is smaller than $0.3$, the missing siblings are added to both sources if the
proximity criterion is met (i.e., if the distance is smaller than half the sum
of the source extents/resolutions). 
If the modulus of the difference between the spectral indices of the two
sources is $\geq 0.3$, the siblings of sources $i$ and $j$ are
modified in the following way: let the number of siblings of source $i$
be greater than that of source $j$. If the numbers of siblings are the same,
let the source with the steeper spectrum be source $i$.
If source $j$ is a sibling of source $i$,
this sibling is removed. In addition, all siblings are removed from source
$j$ that have a different frequency than source $j$. If source $j$ is not
a sibling of source $i$, the common sibling is removed from source $j$.

In a final step the siblings of a given source are compared to the siblings of
all its siblings. If there are less than three common siblings and both spectral
indices are non zero and the numbers of siblings are different, the spectrum of the
source with the smallest number of siblings is removed (i.e. the siblings 
at a different frequency than the source are removed). If this is not the case,
both sources are complemented with the missing siblings.

\subsection{Output \label{sec:output}}

At the end of the data processing for one subfield (RA slices and polar caps), the results are written
in an ASCII file. Since all necessary information for all sources is stored
in the code, the output format can be chosen freely and adapted to the user's purpose. 
For the moment we create two principal outputs: (i) a file with the information necessary
to plot spectra and (ii) a file that can be used as input for SIMBAD.

\section{Code performance \label{sec:performance}}

On a PC with 512~MB RAM and a frequency of 1.4~GHz, the whole data processing of all 3.5 million
sources can be performed in less than 3~h, 
less than one quarter of which is needed to read the input files and to write the final 
ASCII files. The proximity search in one single slice is performed in 5-10~min.
This relatively long time is due to the spherical geometry.

\section{Results \label{sec:results}}

We have based our selection of radio source catalogues on a list of major surveys of discrete radio
sources (see Table~1 of Andernach 1999)\footnote{updated at http://cats.sao.ru/doc/MAJOR\_CATS.html}. 
For the moment we have included the 22 largest of the 66 cited radio catalogues (Table~\ref{tab:entries}). 
These catalogues were downloaded from the VIZIER database. 
\begin{table*}	
      \caption{SPECFIND catalogue entries.}
         \label{tab:entries}
      \[
         \begin{array}{lcrrrrcr}
{\rm name} & {\rm Interferometer\ (I)} & {\rm frequency} & {\rm resolution} & {\rm S}_{\rm min} & {\rm number\ of} & {\rm percentage\ of\ sources\ with} &  {\rm Reference} \\
 & {\rm Single\ Dish\ (S)} & {\rm (MHz)} & {\rm (arcmin)} & {\rm (mJy)} & {\rm sources} & {\rm identified\ spectrum} & \\
\hline
{\rm JVAS} & {\rm I} &  8400 &  5.5\,10^{-3}   & 30 & 2246 & 72 & (1)  \\
{\rm GB6}  & {\rm S} &  4850 &  3.5      &  18 & 75162  & 59 & (2) \\
{\rm 87GB} & {\rm S} &  4850 &  3.5      &  25 & 54579 & 64 & (3) \\
{\rm BWE}  & {\rm S} &  4850 &  3.5      &  25 & 53522  & 61 & (4) \\
{\rm PMN}  & {\rm S} &  4850 &  3.5      &  20 & 50814  & 30 & (5) \\
{\rm MITG} & {\rm S} &  4850 &  2.8      &  40 & 24180 & 52 & (6) \\
{\rm PKS}  & {\rm S} &  2700 &  8.0      &  50 & 8264  & 65 & (7) \\
{\rm F3R} & {\rm S} &  2700 &  4.3      &  40 & 6495 & 61 & (8)  \\
{\rm FIRST} & {\rm I} & 1400 &  0.08333  &  1 & 811117 & 1.5 & (9) \\
{\rm NVSS}  & {\rm I} & 1400 &  0.75     &  2 & 1773484 & 3.6 & (10) \\
{\rm WB}   &  {\rm S} & 1400 & 10.       & 100 &  31524  & 60  & (11) \\
{\rm SUMSS} & {\rm I} &  843  &  0.75     &  8 & 134870 & 1.8 & (12)  \\
{\rm B2}  & {\rm I} &  408  &  8.0       & 250 &  9929 & 72 & (13)  \\
{\rm B3} & {\rm I} & 408  &  5.0      &  100 & 13340 & 66 & (14) \\
{\rm MRC} & {\rm I} & 408 &  3.0      &  700 & 12141 & 73 & (15)     \\
{\rm TXS}  & {\rm I} &  365  &  0.1      &  250 & 66841 & 57 & (16) \\
{\rm WISH} & {\rm I} & 325  &  0.9      &  10 & 90357 & 8.4 & (17) \\
{\rm WENSS} & {\rm I} & 325  &  0.9      &  18 & 229420 & 17 & (18)\\
{\rm MIYUN} & {\rm I} & 232  &  3.8      &  100 & 34426 & 40 & (19) \\
{\rm 4C}  & {\rm I} &  178  &  11.5     &  2000 & 4844 & 53 & (20)  \\
{\rm 3CR} & {\rm I} & 178  &  6.0       &  5000 & 327 & 31 & (21) \\
{\rm 3C}   & {\rm I} & 159  &  10.0      &  7000 & 470 & 3.4 & (22)  \\ 
\end{array}
      \]
\begin{list}{}{}
\item
(1) Patnaik et al. (1992); Browne et al. (1998); Wilkinson et al. (1998)\\
(2) Gregory et al. (1996)\\
(3) Gregory \& Condon (1991)\\ 
(4) Becker et al. (1991)\\
(5) Wright et al. (1994; 1996); Griffith et al. (1994; 1995)\\
(6) Bennett et al. (1986); Langston et al. (1990); Griffith et al. (1990; 1991)\\
(7) Otrupcek \& Wright (1991)\\
(8) F\"{u}rst et al. (1990)\\
(9) White et al. (1998)\\
(10) Condon et al. (1998)\\ 
(11) White \& Becker (1992) \\
(12) Mauch et al. (2003)\\
(13) Colla et al. (1970; 1972; 1973); Fanti et al. (1974)\\
(14) Ficarra et al. (1985)\\
(15) Douglas et al. (1996)\\
(16) Large et al. (1991) \\
(17) de Breuck et al. (2002)\\
(18) Rengelink et al. (1997)\\
(19) Zhang et al. (1997)\\
(20) Pilkington \& Scott (1965); Gower et al. (1967)\\
(21) Bennet (1962)\\
(22) Edge et al. (1959)
\end{list}
\end{table*}
The 3C and 3CR catalogues were added for historical reasons.
The GB6 and 87GB catalogues are not independent. In fact 87GB is based on a subset
of the data used for GB6, which is more sensitive.
The total number of sources is $3\,488\,352$. The NVSS catalogue provides $\sim50$\,\% of these entries.
For this number of sources the sky is divided into 12 RA slices and the 2 polar caps ($|\delta|>70^{\circ}$).

After completion of the data processing, SPECFIND found $757\,894$ independent associations, i.e. sources 
with at least one parent, sibling, or child. 
The number of independent spectra is $66\,866$, of which more than 90\,\% include an NVSS point.
It turns out that SPECFIND identified 5 objects with a spectral index $\alpha < -2$. 
Only pulsars and relic sources in galaxy clusters show such steep spectra. The inspection of these
objects by eye showed that the objects, including a FIRST source, are extended (jet/lobe structure) 
and/or confused.
The ensemble of $66\,866$ independent spectra forms the basis for our further analysis.
The percentage of sources, for which spectra
could be identified by SPECFIND is listed in col.~7 of Table~\ref{tab:entries}.

The highest percentage of sources with detected spectra is 72\% of all sources for the JVAS and B2 surveys.
This is expected for the JVAS, because it is a survey of sources selected on the basis of their 
flat spectral index.  The large surveys produce fewer spectra,
because the other surveys do not cover the same region of the sky as deeply as the former. 
In the SUMSS survey only 1.8\% of the source have spectra, because there is a 
lack of deep enough radio surveys in the southern hemisphere.

\begin{figure}
	\begin{center}
        \resizebox{\hsize}{!}{\includegraphics{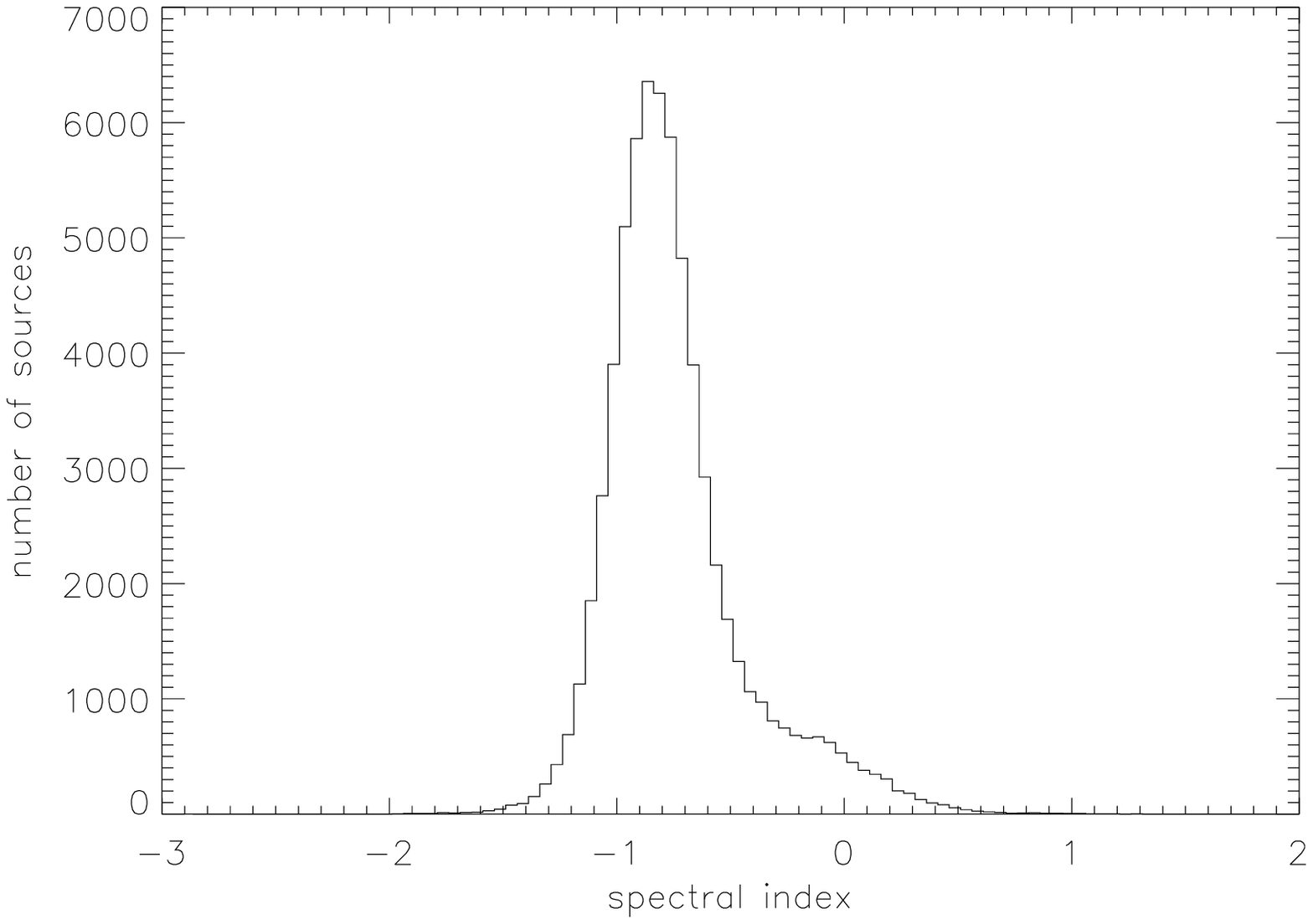}}
	\end{center}
        \caption{Number distribution of spectral indices.
        } \label{fig:result1}
\end{figure}

The distribution of the $66\,866$ spectral indices is shown in Fig.~\ref{fig:result1}.
The distribution peaks at $\alpha \sim -0.9$, which is consistent with previous works
(Vigotti et al. 1989; Kulkarni et al. 1990; Zhang et al. 2003). There is a wing
towards positive spectral indices, which is most probably caused by sources with a flat spectrum 
due to thermal electrons.

\begin{figure}
	\begin{center}
        \resizebox{\hsize}{!}{\includegraphics{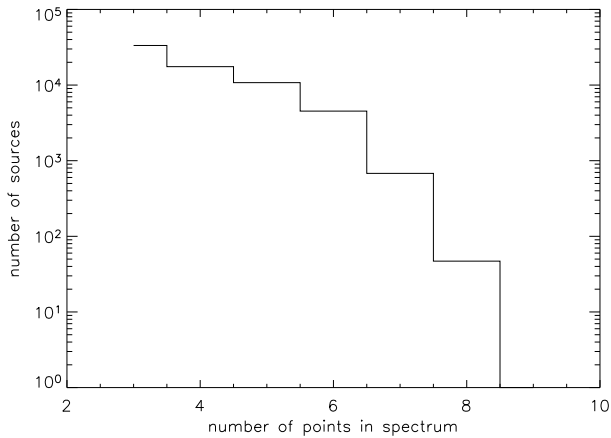}}
	\end{center}
        \caption{The number of sources as a function of the number of independent points in the
	  radio spectra.
        } \label{fig:result2}
\end{figure}

The number of sources as a function of the number of independent frequency points in the
radio spectra for independent sources is shown in Fig.~\ref{fig:result2}.
About $10^{4}$ sources have 5 points and about 50 sources have 8 points. 
The number of sources decreases approximately exponentially between 3 and 6 points
and falls off more rapidly for an even larger number of points.

\begin{figure}
	\begin{center}
        \resizebox{\hsize}{!}{\includegraphics{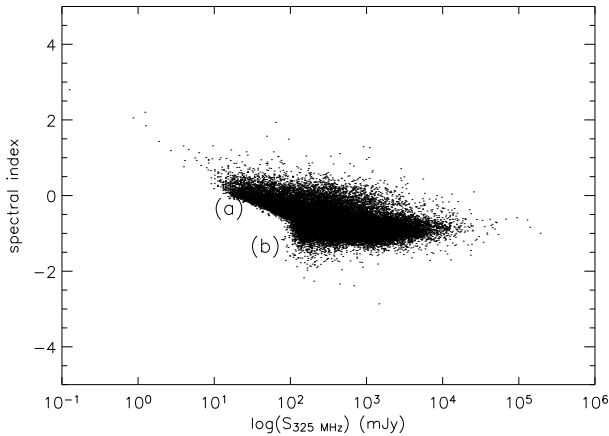}}
	\end{center}
        \caption{The spectral index as a function of the flux density at 325~MHz.
        } \label{fig:result3}
\end{figure}

The spectral index as a function of the flux density at 325~MHz (49~cm; WENSS) is shown in 
Fig.~\ref{fig:result3}.
As already seen in Fig.~\ref{fig:result1}, most of the spectral indices have values around $-0.9$.
Since the NVSS and WENSS surveys are by far the deepest surveys that cover the largest area on the sky
(the whole northern hemisphere), 
the largest number of associations is found for sources in these two catalogues.
Zhang et al. (2003) have correlated them and found $185\,800$ corresponding sources,
which represents the maximum number of sources for which a spectrum can be found.
We found only half of them, because we require at least 3 points for the spectrum.

The straight, almost horizontal edge of the distribution in the left part of the plot 
(marked as (a) in Fig.~\ref{fig:result3}) is due to a selection effect.
For these low flux density sources with a steep spectrum, SPECFIND found a source at 20~cm (NVSS)
and 50~cm (WENSS), but none at 6~cm (GB6, BWE, 87GB, MITG, PMN), where the sensitivity
of the surveys ($\sim 20$~mJy) is insufficient.
At higher flux densities ($S_{325} > 300$~mJy) sources from other
catalogues with lower sensitivities are found. Most of the sources have spectral indices $<-0.6$
(Fig.~\ref{fig:result1}). This translates into a limiting sensitivity of 100~mJy at 325~MHz. 
The number of WENSS sources with flux densities
that exceed this value is $68\,732$, close to the number of identified radio spectra.
The vertical edge in the lower left part of the plot (marked as (b) in Fig.~\ref{fig:result3})
is mainly due to the limiting flux density of the B3 survey. 

If one wants to increase significantly the number of independent
spectra with respect to the existing catalogues, a deep and extended survey 
at an independent wavelength, preferentially smaller than 10~cm, is needed.

\section{Code validation \label{sec:validation}}

In order to validate our algorithm, we compared the spectral indices determined by SPECFIND
with those given in various catalogues.
The radio catalogues in the VIZIER database that include spectral indices are listed in Table~\ref{tab:spectab};
for the PMNS catalogue we calculated the spectral index from the data given
in that catalogue. The columns are: (1) catalogue name,
(2) the frequency used for the cross-identification in SPECFIND, (3) other frequencies for the 
determination of the spectral index, (4) number of sources with spectral index.
The catalogues B3, MITG, PKS, PMNS-S, and FA87 use only two frequencies for the
determination of the spectral index, whereas the other catalogues use more than three
frequencies. The percentage of sources for which a spectrum could be identified
is listed in col. 2 of Table~\ref{tab:standev}. The high percentage of
radio spectra obtained validates the SPECFIND spectrum identification algorithm.
For the direct comparison of spectral indices we only chose indices for which
both methods used approximately the same frequency interval.

Only one pulsar spectrum could be identified by SPECFIND (Table~\ref{tab:pulsars}).
This is due to the small flux densities and steep spectral indices of the pulsar population listed
in PULSARS.
\begin{table}	
      \caption{Pulsar spectrum identified by SPECFIND.}
         \label{tab:pulsars}
      \[
         \begin{array}{lrr}
{\rm source\ name} & {\rm SI\ (SPECFIND)} & {\rm SI\ (PULSARS)} \\
\hline
{\rm B0329+54} & -1.7 & -1.6 \\
\end{array}
      \]
\end{table}

\begin{figure}
\resizebox{\hsize}{!}{\includegraphics{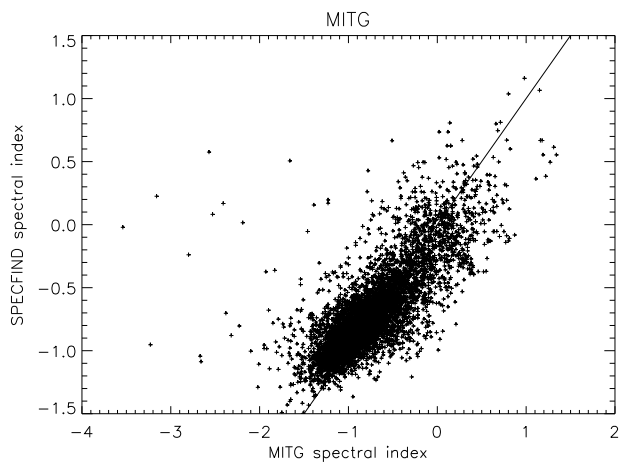}}
\caption{MITG spectral index versus SPECFIND spectral index.} \label{fig:mitg}
\end{figure}

\begin{figure}
\resizebox{\hsize}{!}{\includegraphics{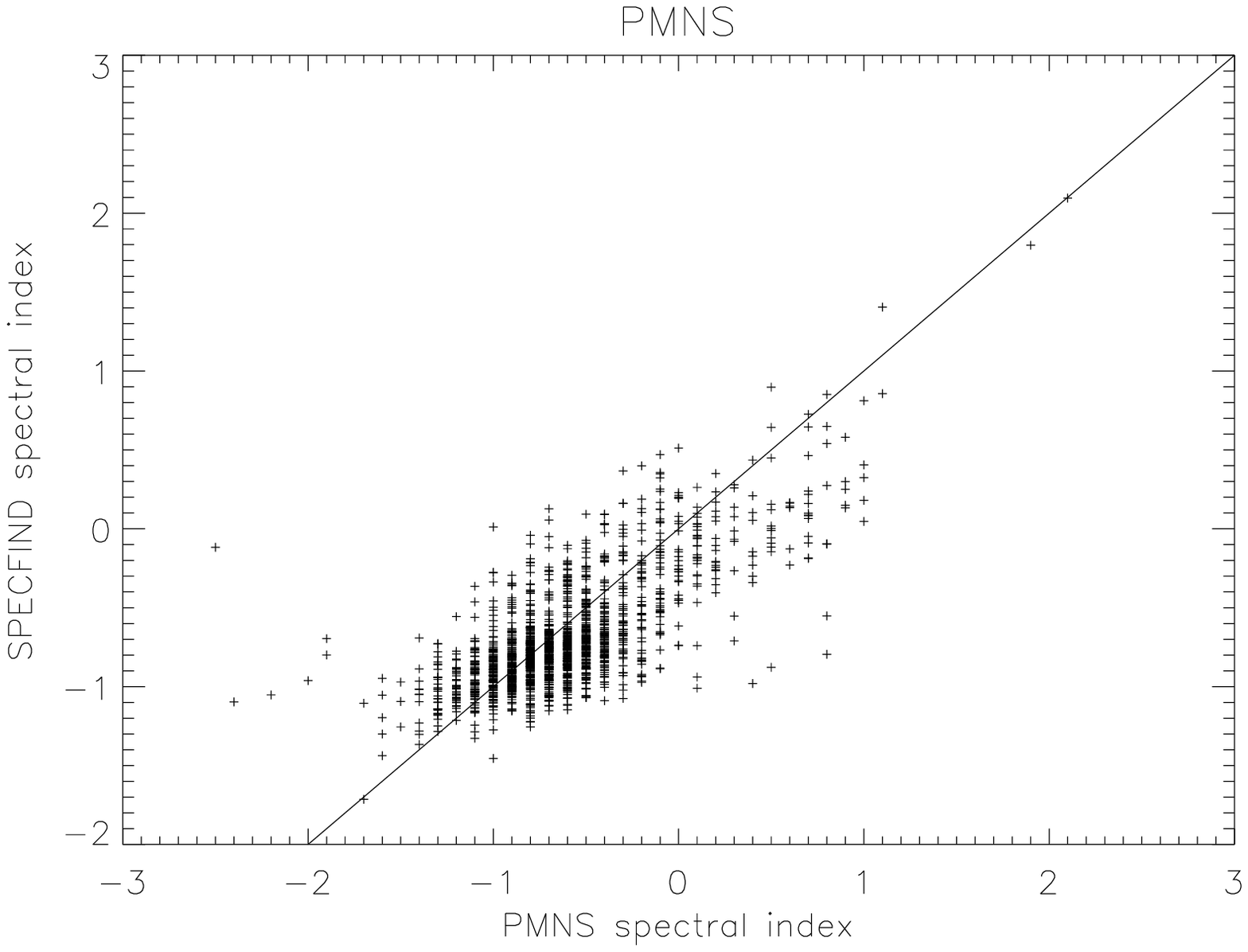}}
\caption{PMN-S spectral index versus SPECFIND spectral index.} \label{fig:pmns}
\end{figure}

\begin{figure}
\resizebox{\hsize}{!}{\includegraphics{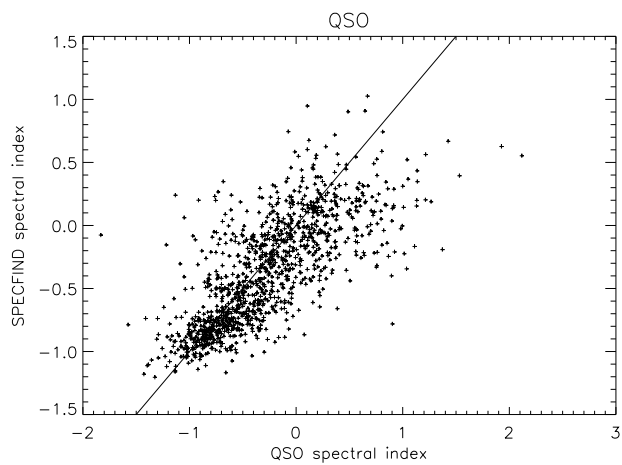}}
\caption{AGN-QSO spectral index versus SPECFIND spectral index.} \label{fig:qso}
\end{figure}

\begin{figure}
\resizebox{\hsize}{!}{\includegraphics{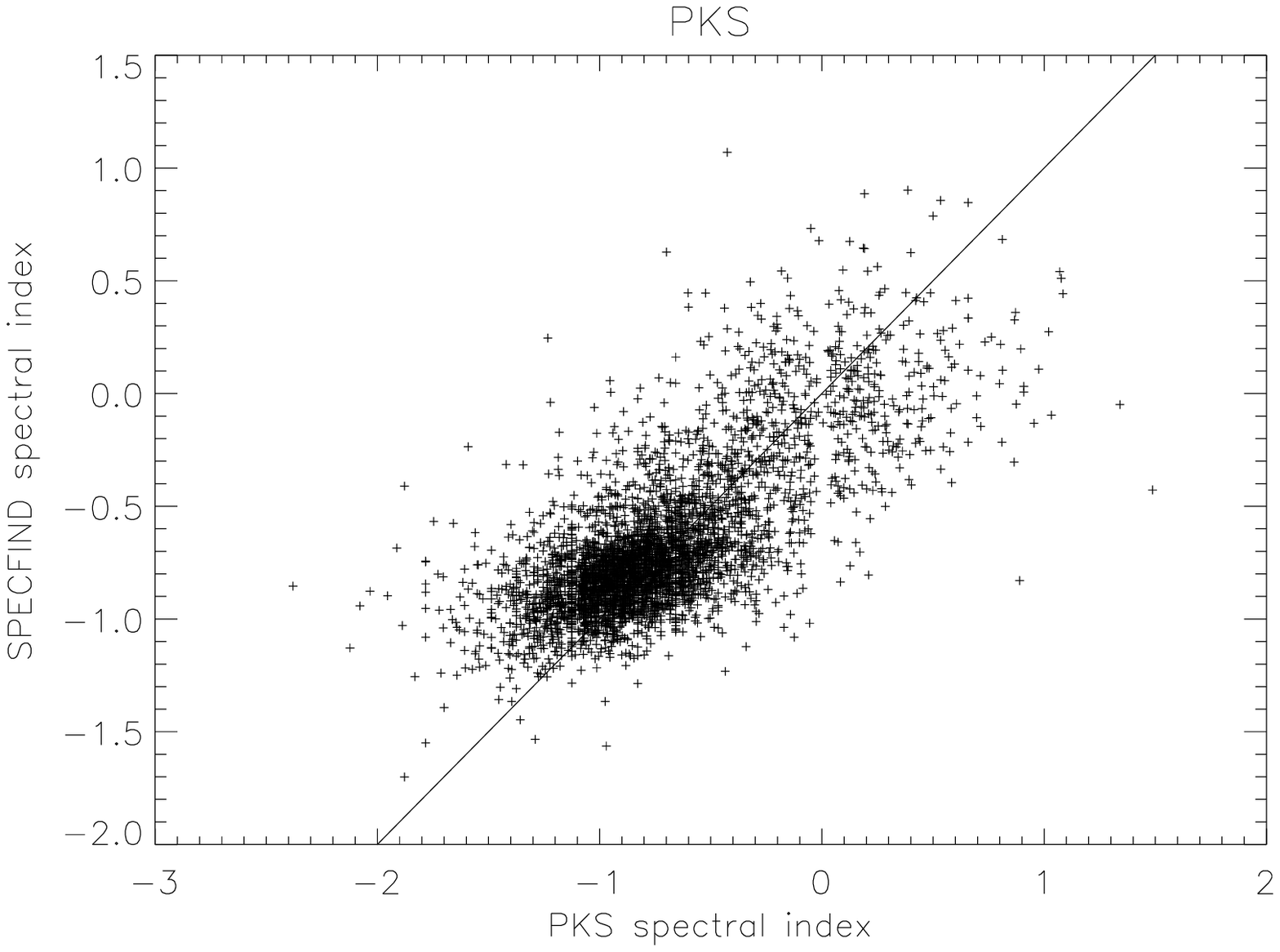}}
\caption{PKS spectral index versus SPECFIND spectral index.} \label{fig:pks}
\end{figure}

\begin{figure}
\resizebox{\hsize}{!}{\includegraphics{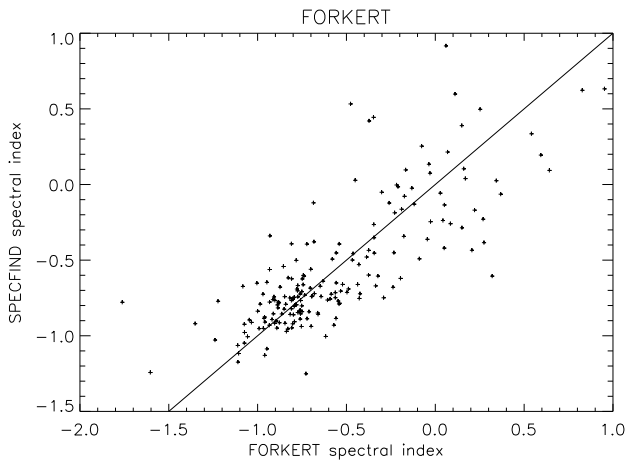}}
\caption{FA87 spectral index versus SPECFIND spectral index.} \label{fig:forkert}
\end{figure}

\begin{figure}
\resizebox{\hsize}{!}{\includegraphics{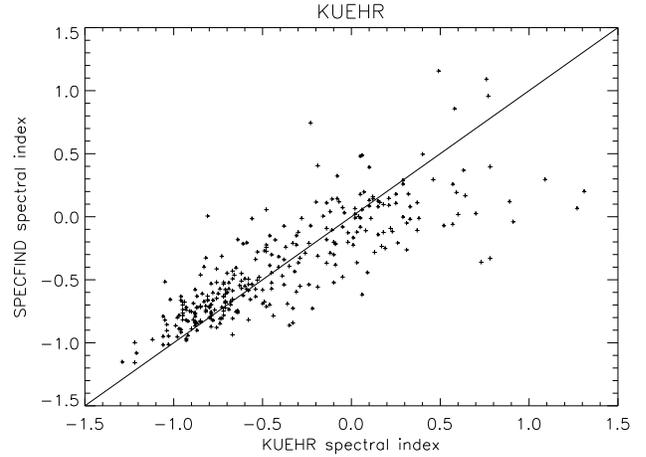}}
\caption{1Jy spectral index versus SPECFIND spectral index.} \label{fig:kuehr}
\end{figure}

\begin{figure}
\resizebox{\hsize}{!}{\includegraphics{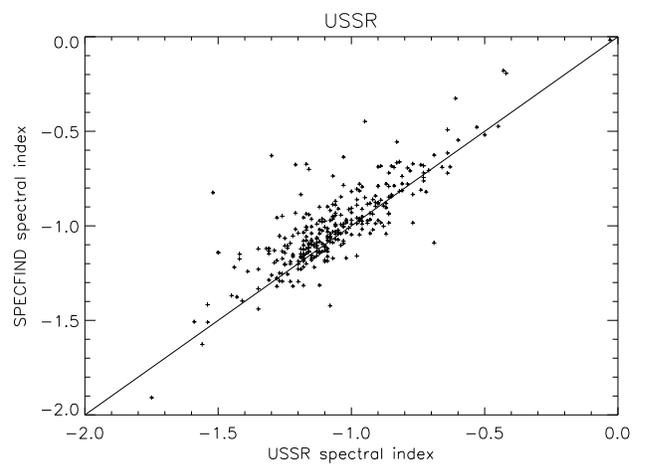}}
\caption{USSR spectral index versus SPECFIND spectral index.} \label{fig:ussr}
\end{figure}

\begin{figure}
\resizebox{\hsize}{!}{\includegraphics{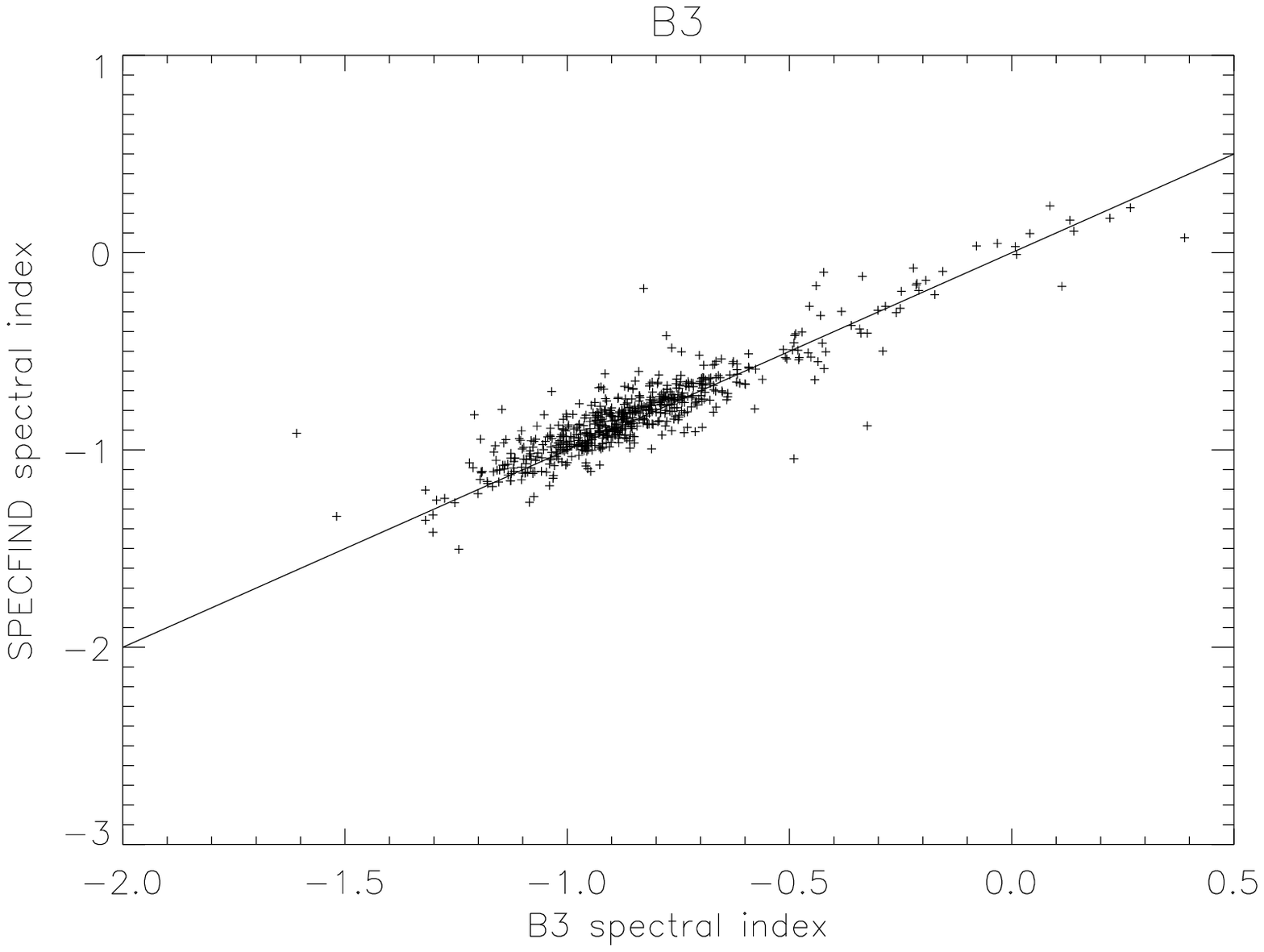}}
\caption{B3 spectral index versus SPECFIND spectral index.} \label{fig:b3}
\end{figure}

A comparison between the spectral indices given in these
catalogues and those found by SPECFIND are shown in Figs.~\ref{fig:mitg}--\ref{fig:b3}.
\begin{table}	
      \caption{Catalogues with spectral indices.}
         \label{tab:spectab}
      \[
         \begin{array}{lrrrr}
{\rm name} & {\rm frequency} & {\rm frequency\ range} &  {\rm number\ of} & {\rm Ref.} \\
 & {\rm (MHz)} & {\rm (MHz)} & {\rm sources} & \\
\hline
{\rm MITG} & 4850 & 365 & 7559 & (1) \\
{\rm PMN-S} & 4850 & 2700 & 1847 & (2) \\
{\rm AGN-QSO} & 4850 & 2700 & 1155 & (3) \\
{\rm PKS} & 2700 & 5000 & 4150 & (4) \\
{\rm FA87} & 2700 & 4750 & 211 & (5) \\
{\rm 1Jy} & 1400 & 2700-5000 & 299 & (6) \\
{\rm USSR} & 1400 & 150-408 & 313 & (7) \\
{\rm PULSARS} & 1400 & 300-1400 & 227 & (8) \\
{\rm B3}  &   408  & 1400 & 613 & (9)
\end{array}
      \]
\begin{list}{}{}
\item
(1) Drinkwater et al. (1997)\\
(2) Wright et al. (1994)\\
(3) V\'{e}ron-Cetty \& V\'{e}ron (2003)\\	
(4) Bennet et al. (1986); Langston et al. (1990); Griffith et al. (1990, 1991)\\
(5) Forkert \& Altschuler (1987)\\
(6) K\"{u}hr et al. (1981)\\
(7) Roettgering et al. (1994)\\
(8) Lorimer et al. (1995); Maron et al. (2000)\\
(9) Kulkarni et al. (1990)\\
\end{list}
\end{table}

In general, the scatter is smaller for smaller spectral indices,
because there is no change in the spectral index in the frequency interval of interest.
The increase of the scatter for large spectral indices (flatter spectra) in the AGN-QSO, FA87 and 1Jy
catalogues (Figs.~\ref{fig:qso}, \ref{fig:forkert}, and \ref{fig:kuehr}) is due to the 
fact that these are radio sources whose spectra peak around 5~GHz. 
The determination of their spectral index therefore depends critically on the
frequency interval used. These intervals are more precisely known for the other catalogues, which
consequently show a smaller scatter. 

The standard deviations $\sigma_{\rm SI}$ of the difference between the spectral indices determined by 
SPECFIND and those determined in the catalogues (Table~\ref{tab:spectab}) are listed in 
Table~\ref{tab:standev}. The best correlation (0.10) between the two is found for the B3 catalogue, and 
the worst (0.36) for the AGN-QSO catalogue. The low consistency of spectral indices in the AGN-QSO
catalogue may be either due to the small frequency range used (4850/2700~MHz), to
variability, or both. In general, the standard deviation is $\sim 0.3$. 
Thus we conclude that the spectral indices determined by SPECFIND have an error of about $\pm 0.3$.
\begin{table}	
      \caption{Standard deviation of the spectral index difference.}
         \label{tab:standev}
      \[
         \begin{array}{lcr}
{\rm catalogue} & {\rm percentage \ of\ sources\ with} & \sigma_{\rm SI} \\
{\rm name}  & {\rm identified\ spectrum} & \\
\hline
{\rm MITG} & 100\% & 0.26 \\
{\rm PMNS-S} & 100\% & 0.29 \\
{\rm AGN-QSO} & 87\% & 0.36 \\
{\rm PKS} & 100\% & 0.30 \\
{\rm FA87} & 91\% & 0.28 \\
{\rm 1Jy} & 100\% & 0.29 \\
{\rm USSR} & 99\% & 0.13 \\
{\rm PULSARS} & - & - \\
{\rm B3}  & 100\% & 0.10 \\
\end{array}
      \]
\end{table}

\section{Conclusions \label{sec:conclusions}}

SPECFIND is a very efficient tool to identify radio spectra using radio catalogues
of different formats. SPECFIND can handle radio surveys of very different resolutions and
sensitivities. It has been applied to 22 survey catalogues at 11 different frequencies containing 
a total of 3.5 million sources, leading to more than $700\,000$ independent radio cross-identifications
and $\sim 67\,000$ independent radio spectra with more than two independent frequencies.
The code was tested and its results validated by a comparison between the spectral indices found by SPECFIND and
those determined by other authors. The determined spectral indices have an error of about $\pm0.3$. 
Negative spectral indices have smaller errors, while the error of positive spectral
indices can be larger, mainly because of the occurrence of a peak in the spectrum.
The code is quite rapid (less than 3~hr running time on a standard PC for 3.5 million sources) and since it is 
written in C, it can be run on virtually all PCs with at least 512~MB RAM. It produces an output 
of variable format that can be adapted easily to the purpose of the user. 
A special output to enter the cross-identifications into SIMBAD has been developed.
The code has been optimised to find a maximum number of spectra with a relatively small number 
of misidentifications. It represents thus a promising tool to extract radio spectra from a large 
sample of radio catalogues, like those stored in VIZIER. 
Although at present only 22 catalogue entries can be accessed  
by SPECFIND, this number will be increased in the future.
The advantage of this procedure is that radio spectra and cross-identifications are established
in advance, and can be stored in a 'master list' within VIZIER, allowing a fast and efficient access.
We expect to make the cross-identifications soon available via the VIZIER database.
 
\begin{acknowledgements}
We would like to thank H.~Andernach for very helpful discussions and O.V.~Verkhodanov for
his comment on ambiguous sources and ``centers of mass''.
\end{acknowledgements}

\end{document}